\documentclass[graybox]{svmult}

\usepackage{helvet}         
\usepackage{courier}        
\usepackage{type1cm}        
\usepackage{makeidx}         
\usepackage{graphicx}        
\usepackage{multicol}        
\usepackage[bottom]{footmisc}

\usepackage{hyperref}

\usepackage{amsmath}
\usepackage{dsfont}
\usepackage{amsfonts}
\usepackage{array}

\def\be{\begin{equation}}
\def\ee{\end{equation}}
\def\bes{\begin{subequations}}
\def\ees{\end{subequations}}

\def\i{{\rm i}}

\def\d{{\rm d}}
\def\ul{\underline}

\def\nn{{\nonumber}}
\def\q{{\mathfrak{q}}}
\def\t{{\mathfrak{t}}}
\def\p{{\mathfrak{p}}}
\def\L{{\textrm{L}}}
\def\R{{\textrm{R}}}
\newcommand{\mb}[1]{\mathbf{#1}}

\def\E{{\mb E}}

\begin{document}

\title*{Defects at the Intersection: the Supergroup Side}
\author{Fabrizio Nieri}
\institute{Fabrizio Nieri \at Hamilton Mathematics Institute, Trinity College Dublin (ROI) \& DESY (Germany), \email{fb.nieri@gmail.com}}
\maketitle

\abstract{We consider two seemingly different theories in the $\Omega$-background: one arises upon the most generic Higgsing of a 5d $\mathcal{N}=1$ $\text{U}(N)$ gauge theory coupled to matter, yielding a 3d-1d intersecting defect; the other one arises upon simple Higgsing of a 5d $\mathcal{N}=1$ $\text{U}(N|M)$ supergroup gauge theory coupled to super-matter, yielding another defect. The cases $N=M=1$ are discussed in detail via equivariant localization to matrix-like models. The first theory exhibits itself a supergroup-like structure, which can be motivated via non-perturbative string dualities, and in a matter decoupling limit it is argued to be dual to a supergroup version of refined Chern-Simons theory. Furthermore, it is observed that the partition functions of the two defect theories are related by analytic continuation in one of the equivariant parameters. We find a common origin in the algebraic engineering through $q$-Virasoro screening currents. Another simple Higgsing of the 5d $\mathcal{N}=1$ $\text{U}(1|1)$ yields a single component defect whose partition function is reminiscent of ordinary refined Chern-Simons on a lens space.}

\section{Intersecting Defects and Supermatrix-like Models}

Gauge invariance is one of the main principles behind our comprehension of Nature, and the dichotomy between matter particles and force mediators (fermions vs. bosons) in theories based on ordinary compact Lie groups is a fact of life. Supergroups, on the other hand, unify particles of opposite statistics: while theories exhibiting global supergroup symmetries have been long appreciated (cf. SUSY models), it is also quite interesting to study QFT (supersymmetric or not) based on gauge supergroups. Non-unitarity is manifest due to the violation of spin-statistics, whereas the lack of a definite bilinear form on the  gauge algebra requires an intrinsic non-perturbative approach. These are actually some of the features which make such theories and the closely related supermatrix models interesting to investigate. In fact, they do arise, in one way or another, in many places of theoretical physics: effective membrane dynamics \cite{Aharony:2008ug,Aharony:2008gk,Kapustin:2009kz,Drukker:2009hy}, analytic continuations of unitary models \cite{Marino:2009jd,Awata:2012jb}, topological strings \cite{Drukker:2010nc,Drukker:2011zy,Marino:2011eh,Hatsuda:2013oxa}, exotic phenomena \cite{Vafa:2014iua,Dijkgraaf:2016lym}, instanton calculus  \cite{Kimura:2019msw} and integrability \cite{Chen:2019vvt,Chen:2020rxu,Nekrasov:2018gne} just to mention few examples. 

\begin{figure}[!ht]
\begin{center}
\includegraphics[width=0.45\textwidth]{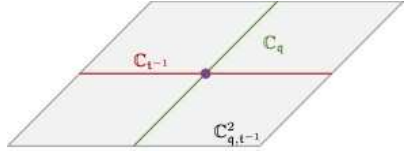}
\end{center}
\vspace*{-10pt}
\caption{Representation of the supports of the 5d theory and of the 3d-1d defects. At each point there is also a circle which is not explicitly displayed.}
\label{fig:C2omega}
\end{figure}

This note,
a brief account of the results published in \cite{Kimura_2021} (supplemented by original computations in Subsection \ref{subMatter}), is about yet another place where supergroup gauge theories, a supergroup version of refined Chern-Simons theory in particular \cite{Aganagic:2011sg} (see also \cite{Cassia:2021uly} for recent work in this direction), show up: SUSY theories based on ordinary gauge groups  but supported on intersecting subspaces embedded in an ambient space. The intersecting gauge theories of our interest arise upon Higgsing a parent 5d SUSY gauge theory with unitary group in the $\Omega$-background $\mathbb{C}^2_{\q,\t^{-1}}\times\mathbb{S}^1$, while the support of the defects is given by the two orthogonal cigars $\mathbb{C}_\q\times \mathbb{S}^1$ and $\mathbb{C}_{\t^{-1}}\times \mathbb{S}^1$ intersecting at the origin along a common circle (Fig. \ref{fig:C2omega}).
\vspace*{-8pt}
\vspace*{-8pt}

This description of the 3d-1d coupled system allows its partition function to be computed by specialization of the instanton partition function of the parent 5d theory, which for the $\text{U}(N)$ SQCD can be presented in a combinatorial form as a summation over a set of $N$ integer partitions $\lambda$ \cite{Nekrasov:2002qd,Nekrasov:2003rj}
\begin{align}
Z_\text{inst.}[\text{SQCD}]\equiv & \ \sum_{\{\lambda_A\}}\Lambda^{\sum_A|\lambda_A|}\, Z_{\{\lambda_A\}}[\text{SQCD}]\xrightarrow{\nu\to \nu^*} Z_\text{vortex}[\textrm{Defect}_{\q,\t}]~,\nn\\
Z_{\{\lambda_A\}}[\text{SQCD}]\equiv & \ \prod_{A,B=1}^N\frac{N_{\emptyset \lambda_A}(\nu_A/\bar \mu_B;\q,\t)N_{\lambda_A \emptyset}(\mu_B/\nu_A;\q,\t)}{N_{\lambda_A \lambda_B}(\nu_A/\nu_B;\q,\t)}~,\label{eq:ZSQCD}
\end{align}
where $\Lambda$ denotes the instanton counting parameter, $\mu,\bar\mu$ (anti-)fundamental flavor fugacities and $\nu$ the Coulomb branch parameters (our conventions and definitions are set in the Appendix). The truncation to the vortex part of the defect partition function is achieved by locking the Coulomb  parameters to the flavour ones as follows
\be
\nu_A\to \nu_A^*\equiv \mu_A \t^{-r_A}\q^{c_A}~,\quad r,c\in\mathbb{Z}_{\geq 0}^N~.
\ee
The defect theory is UV described by a pair of 3d $\mathcal{N}=2$ gauge theories with gauge groups $\text{U}(r)$ and $\text{U}(c)$ respectively (coupled to adjoint and fundamental/anti-fundamental chirals) and interacting through 1d chiral matter along the common $\mathbb{S}^1$ at the origin, supplemented with superpotential terms (enforcing identifications among parameters) \cite{Gomis:2016ljm,Pan:2016fbl,Nieri:2017ntx,Nieri:2018pev}. The analysis of the partition function, which can be recast in a matrix-like integral
\begin{multline}\label{eq:DefectMMqtinv}
Z[\text{Defect}_{\q,\t}]\equiv\oint\prod_{a=1}^r\frac{\d z_a^\R}{2\pi\i z_a^\R}\, \prod_{b=1}^c\frac{\d z_b^\L}{2\pi\i z_b^\L}\,\Big(\prod_{a}  z_a^\R\Big)^{-\zeta_\R}\, \Big(\prod_{b}  z_b^\L\Big)^{\zeta_\L}\times\\
\times \Delta(z^\R,z^\L;\q,\t^{-1},\p)\times  \prod_{A,a}\frac{(z^\R_a/\eta_\R \bar\mu_A;\q)_\infty}{( z^\R_a/\eta_\R\mu_A;\q)_\infty}\prod_{A,a}\frac{( z^\L_a/\eta_\L\bar\mu_A;\t^{-1})_\infty}{( z^\L_a/\eta_\L\mu_A;\t^{-1})_\infty}~,
\end{multline}
where $\p^{1/2}\equiv \sqrt{\q\t^{-1}}$, $\eta_\R/\eta_\L\equiv1/\sqrt{\q\t}$, $\q^{\zeta_\R}\equiv \Lambda\equiv \t^{\zeta_\L}$, suggests that the intersecting defects provide a deformation of a dual supergroup gauge theory, which for $N=1$ (SDEQ) can be identified as a supergroup version of refined Chern-Simons theory on $\mathbb{S}^3$  (after a matter decoupling limit).
\vspace*{-8pt}
\begin{figure}[!ht]
\begin{center}
\includegraphics[width=0.8\textwidth]{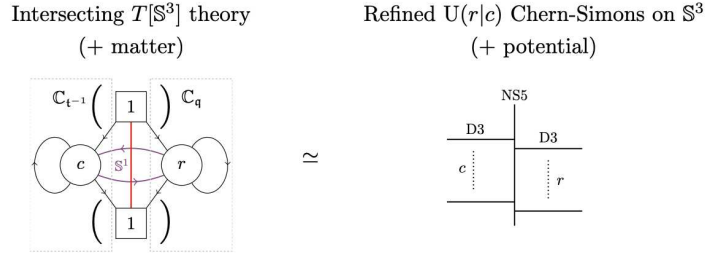}
\end{center}
\vspace*{-10pt}
\caption{The 3d[intersecting]-3d[supergroup] equivalence, here specialized to $N=1$. It can be placed in the context of a generalized 3d-3d correspondence. The squares on the l.h.s. represent coupling to (anti-)fundamental matter (which can be decoupled).}
\label{fig:3d3dD}
\end{figure}
\vspace*{-8pt}
This is essentially because the integration measure features the supergroup version of the Macdonald weight function \cite{sergeev2008deformed,atai2021supermacdonald}
\bes
\begin{align}\label{eq:superMacK}
\Delta(z^\R,z^\L;\q,\t^{-1},\p)\equiv & \ \frac{\Delta_\t(\ul z^\R;\q)\Delta_{\q^{-1}}(\ul z^\L;\t^{-1})}{\prod_{a,b}(1-\p^{-1/2} z^\L_b/z^\R_{a})(1-\p^{-1/2}z^\R_a/z^\L_{b})}~,\\
\Delta_\t(z;\q)\equiv & \ \prod_{i\neq j}\frac{(z_i/z_j;\q)_\infty}{(\t z_i/z_j;\q)_\infty}~,
\end{align}
\ees
which in scaling limit $\q,\t\to1$ ($\ln\t/\ln\q\equiv \beta$ fixed) yields the $\beta$-deformation of the Cauchy weight function \cite{Atai_2019}, simply related to the Hermitian supermatrix measure in the unrefined case $\beta=1$.\footnote{Let us note that, because of the original $\q\leftrightarrow \t^{-1}$ symmetry, we considered here the chamber $|\q|<1,|\t^{-1}|<1$, which is, however, tricky for the unrefined limit. The case $|\q|<1,|\t|<1$ is discussed in \cite{Kimura_2021}.} The democracy between the two orthogonal planes opens up the possibility of understanding the refinement $\t\neq \q$ as a specific deformation away from the supergroup point: $\ln\q$ and $\ln\t^{-1}$ are identified with the inverse Chern-Simons couplings, strictly opposite for a supergroup theory. This can roughly be seen in the decoupling limit $\mu\to 0,\bar\mu\to \infty$, in which case the potential asymptotically contributes with $\exp[\sum_a(\ln z^\R_a)^2/\ln\q-\sum_b(\ln z^\L_b)^2/\ln\t]$. The appearance of a supergroup structure may seem surprising from the intersecting defect perspective: the parent theory is an ordinary gauge theory, while after Higgsing the  degrees of freedom are even defined on different space-time components. An explanation can be offered by embedding the configuration under study into string/M-theory and then exploiting chain of non-perturbative dualities to recognise a setup engineering supergroup Chern-Simons theory \cite{Vafa:2001qf,Okuda:2006fb,Mikhaylov:2014aoa} (Fig. \ref{fig:3d3dD}). 

\section{Towards Non-Unitary Open/Closed Duality}

The relation between the 5d parent theory and the 3d defect in the ordinary Higgsing process (i.e. $c=0$), once embedded into string theory, can also be understood as a large $r$ geometric transition (open/closed duality) \cite{Gopakumar:1998ki}. Since in a generic Higgsing process the parent 5d theory (the closed side) gives rise to both $\q$- and $\t$-branes supporting the intersecting defect in space-time (the open side), it is natural to wonder what happens upon trying to go back by a simultaneous large $(r,c)$ limit. One may think that the original setup must be recovered, however, this is not the only possibility: it turns out that a closed string side engineering a 5d supergroup gauge theory \cite{Kimura:2019msw} (see \cite{Kimura:2020jxl} for a nice review) is also possible (and more natural to some extent).  

A clean illustration of such dynamics can be achieved by recalling that the toric geometries/brane configurations dual to our 5d-3d-1d theories can be thought of as networks of Ding-Iohara-Miki (DIM) intertwiners \cite{Awata:2016riz,Zenkevich:2018fzl}. In the special case we are interested in (SQCD or SQED) the algebraic description can also be recast in terms of the $q$-Virasoro algebra \cite{Shiraishi:1995rp,frenkelR}, whose screening currents $\mb S^\pm$ are identified with $\q$- and $\t$-branes. In the ordinary Higgsing only one type of screening is considered, then the open/closed duality is the observation that free field correlators involving a finite number of charges capture 3d defect partition functions \cite{Aganagic:2013tta}, whereas sending their number to infinity \cite{Kimura:2015rgi} recovers the partition function of the parent 5d theory\footnote{The dependence on vortex or instanton counting parameters arises from the zero mode contributions acting on the charged Fock vacuum.}
\begin{multline}
Z[\text{Defect}_\q]\simeq \langle\oint \prod_{i=1,\ldots,r}^{\prec} \d z_i\, \mb S^+(z_i)\ \cdots \rangle \xrightarrow{r \to \infty}\\
\xrightarrow{r \to \infty}  \langle \sum_{\{k_i\in\mathbb{Z}\}}\prod^{\prec}_{i\geq 1}x_i\q^{k_i}\mb S^+(x_i\q^{k_i}) \ \cdots \rangle \simeq Z_\text{inst.}[\text{U}(1)]~,
\end{multline}
where the infinitely-many base points in the Jackson integration are at $x_i\equiv \eta_+\nu_+ \t^{i-1}$, for some $\eta_+\in\mathbb{C}^\times$.\footnote{We are focusing on the adjoint sector only as the addition of fundamental matter can be implemented by inserting vertex operators, represented by the dots. Also, $\prec$ means the product runs in increasing order from left to right, and viceversa for $\succ$.} For simplicity, we are here limiting to the Abelian theory in 5d (dual to the resolved conifold geometry). Both the 3d integrand and the 5d instanton summands are easily recognized from the OPE relations 
\bes
\begin{align}
\mb{S}^{(\mp)}(x)\mb{S}^{(\pm)}(x')=& \  \ :\mb{S}^{(\mp)}(x)\mb{S}^{(\pm)}(x'): \ \frac{(-\p^{1/2}x x')^{-1}}{(1-\p^{-1/2}x/x')(1-\p^{-1/2}x'/x)}~,\\
\mb{S}^{(+)}(x)\mb{S}^{(+)}(x')=& \ :\mb{S}^{(+)}(x)\mb{S}^{(+)}(x'):\ \frac{(x'/x;\q)_\infty (\p x'/x;\q)_\infty}{(\q x'/x;\q)_\infty(\t x'/x;\q)_\infty}\, x^{2\beta}~,\\
\mb{S}^{(-)}(x)\mb{S}^{(-)}(x')=& \ :\mb{S}^{(-)}(x)\mb{S}^{(-)}(x'):\ \frac{(x'/x;\t)_\infty (\p^{-1} x'/x;\t)_\infty}{(\t x'/x;\t)_\infty(\q x'/x;\t)_\infty}x^{2\beta^{-1}}~.
\end{align}
\ees
where $:~:$ denotes the usual (free boson) normal ordering. The characteristic summation over partitions arises due to zeros in the coefficients for configurations of points outside of the set
\be
\chi^+\equiv \{\nu_+\t^{i-1}\q^{-\lambda^+_i},\lambda^+_i\geq \lambda^+_{i+1},\i\in[1,+\infty)\}~.
\ee

Mimicking the same logic, we can consider free field correlators involving both types of screenings. When there are $r$ of one type and $c$ of the other type, the integration measure of the 3d-1d intersecting defect partition function is manifestly reproduced as in (\ref{eq:superMacK}).\footnote{Up to overall $\q$- and $\t$-constant which play little role for the identification.} When an infinite amount of both types is considered, we first generate a second dynamical set
\be
\widetilde\chi^-\equiv \{\nu_-\q^{i-1}\t^{-\lambda^{-\vee}_i},\lambda^{-\vee}_i\geq \lambda^{-\vee}_{i+1},\i\in[1,+\infty)\}~,
\ee
where ${}^\vee$ denotes transposition, and it is convenient to fix $\eta_+/\eta_-\equiv \p^{1/2}$.\footnote{Let us note that because of the $\q\leftrightarrow \t$ exchange symmetry of the points in the two sets, we are naturally led to consider the chamber $|\q|,|\t|<1$.} The diagonal (i.e. $++$ and $--$) OPE factors simply generate the adjoint instanton summands for the $\text{U}(1)\times\text{U}(1)$ theory
\bes
\begin{align}
&\prod_{x\in\chi^+}^\prec \mb S^+(\eta_+x)\simeq  \prod_{i,j=1}^{\ell(\lambda^+)}\frac{(\t\, \t^{j-i};\q)_\infty}{( \t^{j-i};\q)_\infty}\frac{(\t^{j-i}\q^{\lambda^+_i-\lambda^+_j};\q)_\infty}{(\t\, \t^{j-i}\q^{\lambda^+_i-\lambda^+_j};\q)_\infty}\times\nn\\
&\times\frac{1}{N_{\lambda^+ \emptyset}(\t^{\ell(\lambda^+)};\q,\t)N_{\emptyset \lambda^+}(\t^{-\ell(\lambda^+)};\q,\t)}= \frac{1}{N_{\lambda^+\lambda^+}(1;\q,\t)}\equiv Z_{++}^\text{v}~,\\
& \prod_{x\in\widetilde\chi^-}^\succ \mb S^-(\eta_-x)\simeq \prod_{i,j=1}^{\ell(\lambda^{-\vee})}\frac{(\q\,\q^{j-i};\t)_\infty}{(\q^{j-i};\t)_\infty}\frac{(\q^{j-i}\t^{\lambda^{-\vee}_i-\lambda^{-\vee}_j};\t)_\infty}{(\q\, \q^{j-i}\t^{\lambda^{-\vee}_i-\lambda^{-\vee}_j};\t)_\infty}\times\nn\\
&\times\frac{1}{N_{\lambda^{-\vee} \emptyset}(\q^{\ell(\lambda^{-\vee})};\t,\q)N_{\emptyset \lambda^{-\vee}}(\q^{-\ell(\lambda^{-\vee})};\t,\q)}\!=\! \frac{1}{N_{\lambda^{-\vee}\lambda^{-\vee}}(1;\t,\q)}\!\equiv\! \p^{-|\lambda^-|}Z_{--}^\text{v}~,
\end{align}
\ees
where the equalities are up to  normalization (empty diagrams) and zero mode contributions. This is the first sign of a super-instanton expansion, however, the crucial information comes from the mixed (i.e. $+-$ and $-+$) terms
which have to match the specific bi-fundamental-like contributions. This can checked, for instance, by using the identities 
\bes
\begin{align}
&\frac{\prod_{(x,x')\in \widetilde\chi^-_\emptyset\times\chi^+_\emptyset}
 (1-\p^{-1/2}\eta_-\eta_+^{-1}x/x')}{\prod_{(x,x')\in \widetilde\chi^-\times\chi^+} (1-\p^{-1/2}\eta_-\eta_+^{-1}x/x')}= \nn\\
  =& \  \frac{\prod_{i=1}^{\ell(\lambda^+)}\prod_{j=1}^{\ell(\lambda^{-\vee})}(1-\p^{-1}\nu_-/\nu_+ \t^{-i}\q^{j})}{\prod_{i=1}^{\ell(\lambda^+)}\prod_{j=1}^{\ell(\lambda^{-\vee})}(1-\p^{-1}\nu_-/\nu_+ \t^{-\lambda^{-\vee}_j-i}\q^{\lambda^{+}_i+j})}\times\nn\\
  \times & \ N_{\lambda_+\emptyset}(\q^{\ell(\lambda^{-\vee})}\nu_-/\nu_+;\q,\t)N_{\emptyset\lambda_-^\vee}(\p^{-1}\t^{-\ell(\lambda^{+})}\nu_-/\nu_+;\t,\q)~,\\
&\frac{\prod_{(x,x')\in \widetilde\chi^-_\emptyset\times\chi^+_\emptyset}
 (1-\p^{-1/2}\eta_+\eta_-^{-1}x'/x)}{\prod_{(x,x')\in \widetilde\chi^-\times\chi^+} (1-\p^{-1/2}\eta_+\eta_-^{-1}x'/x)}=\nn\\
 = & \ \frac{\prod_{i=1}^{\ell(\lambda^+)}\prod_{j=1}^{\ell(\lambda^{-\vee})}(1-\nu_+/\nu_- \t^{i}\q^{-j})}{\prod_{i=1}^{\ell(\lambda^+)}\prod_{j=1}^{\ell(\lambda^{-\vee})}(1-\nu_+/\nu_- \t^{\lambda^{-\vee}_j+i}\q^{-\lambda^+_i-j})}\times\nn\\
 \times& \ N_{\emptyset\lambda_+}(\q^{-\ell(\lambda^{-\vee})}\nu_+/\nu_-;\q,\t)N_{\lambda_-^\vee\emptyset}(\p^{-1}\t^{\ell(\lambda^{+})}\nu_+/\nu_-;\t,\q)~.
\end{align}
\ees
Once all the pieces are combined, the super-instanton partition function of the $\text{U}(1|1)$ theory is reproduced. In particular, the counting parameters  $\Lambda^{\pm 1}$ in the $\pm$ sectors are accounted by the zero modes of the screening currents.

\subsection{Inclusion of Matter and Supergroup Higgsings}\label{subMatter}

The inclusion of (anti-)fundamental matter with supergroup flavor symmetry introduces the following contributions to the super-instanton summands
\bes
\begin{align}
Z^\text{f}_{+\pm}\equiv & \ N_{\emptyset\lambda^+}(\nu_+/\bar\mu_\pm;\q,\t)^{\pm 1}~,& Z^\text{af}_{+\pm}\equiv & \ N_{\lambda^+\emptyset}(\mu_\pm/\nu_+;\q,\t)^{\pm 1}~\\
Z^\text{f}_{-\pm}\equiv & \ N_{\emptyset\lambda^{-\vee}}(\p^{-1}\nu_-/\bar\mu_\pm;\t,\q)^{\mp 1}~,& Z^\text{af}_{-\pm}\equiv & \ N_{\lambda^{-\vee}\emptyset}(\p^{-1}\mu_\pm/\nu_-;\t,\q)^{\mp 1}~.
\end{align}
\ees
In particular, the diagonal sectors show zeros at particular values of $\nu_\pm$ which can be used to truncate the partition function.\footnote{For defects induced by orbifolding see \cite{Chen:2020rxu}.} Let us consider the inclusion of one anti-fundamental, then the interesting points are at
\be
\mu_+/\nu_+=\t^r\q^{-c'}~,\quad 
\mu_-/\nu_-=\p\q^c\t^{-r'}~,\quad r,c,r',c'\in\mathbb{Z}_{\geq0}~.
\ee
Viewing the $\text{U}(1)$ theory as the subsector $\text{U}(1|0)$, we see that the situation described in the previous Section corresponds to $c=r'=0$, which completely freezes the negative node. With an abuse of notation, this defect can be dubbed $\text{U}(r|0)\times\text{U}(0|c')$, while the most general one as $\text{U}(r|c)\times\text{U}(r'|c')$. In the following, we consider a couple of intermediate possibilities.\\

\noindent\textbf{Intersecting $\text{U}(r|c)$ Defect.} We are here interested in setting
\be
\nu_+\to \nu_+^*\equiv \mu_+\t^{-r}~,\quad \nu_-\to \nu_-^*\equiv \mu_-\p^{-1}\q^{-c}~,
\ee
so that $\ell(\lambda^+)\leq r$, $\ell(\lambda^{-\vee})\leq c$. At these points, the diagonal vector and fundamental contributions partially simplify
\bes
\begin{align}
Z^\text{v}_{++}Z^\text{af}_{++}\xrightarrow{\nu_+\to \nu_+^*}& \ \prod_{i,j=1}^{r}\frac{(\t\,\t^{j-i};\q)_\infty}{(\t^{j-i};\q)_\infty}\frac{(\t^{j-i}\q^{\lambda^+_i-\lambda^+_j};\q)_\infty}{(\t\, \t^{j-i}\q^{\lambda^+_i-\lambda^+_j};\q)_\infty}\frac{1}{N_{\emptyset \lambda^+}(\t^{-r};\q,\t)}~,\\
Z^\text{v}_{--}Z^\text{af}_{--}\xrightarrow{\nu_-\to \nu_-^*}& \ \prod_{i,j=1}^{c}\frac{(\q\,\q^{j-i};\t)_\infty}{(\q^{j-i};\t)_\infty}\frac{(\q^{j-i}\t^{\lambda^{-\vee}_i-\lambda^{-\vee}_j};\t)_\infty}{(\q\, \q^{j-i}\t^{\lambda^{-\vee}_i-\lambda^{-\vee}_j};\t)_\infty}\frac{\p^{|\lambda_-|}}{N_{\emptyset \lambda^{-\vee}}(\q^{-c};\t,\q)}~,
\end{align}
\ees
and the mixed contributions too
\begin{multline}
Z^\text{v}_{+-}Z^\text{af}_{+-}Z^\text{v}_{-+}Z^\text{af}_{-+} \xrightarrow{\nu_\pm\to \nu^*_\pm}   \prod_{i=1}^{r}\prod_{j=1}^{c}\frac{1-\p\mu_+/\mu_- \t^{i-r}\q^{c-j}}{1-\p\mu_+/\mu_- \t^{\lambda^{-\vee}_j+i-r}\q^{-\lambda^+_i+c-j}}\times\\
\times\prod_{i=1}^{r}\prod_{j=1}^{c}\frac{1-\p^{-2}\mu_-/\mu_+ \t^{r-i}\q^{j-c}}{1-\p^{-2}\mu_-/\mu_+ \t^{-\lambda^{-\vee}_j+r-i}\q^{\lambda^{+}_i+j-c}}\times\\
\times \p^{-|\lambda^+|}N_{\emptyset\lambda_+}(\p^2\t^{-r}\mu_+/\mu_-;\q,\t) N_{\emptyset\lambda_-^\vee}(\p^{-2}\q^{-c}\mu_-/\mu_+;\t,\q)~.
\end{multline}
Eventually, the truncated super-instanton summands can be organized as the (normalized) residues at the poles
\be
z^+_i=\alpha_+\mu_+ \t^{i-r}\q^{-\lambda_i^+}~,\quad z^-_j=\alpha_-\mu_- \q^{j-c}\t^{-\lambda_j^{-\vee}}~
\ee
in the contour integral 
\begin{multline}
Z_\text{inst.}[\text{U}(1|1)+\text{af}]\xrightarrow{\nu_\pm\to \nu_\pm^*}\oint\prod_{i=1}^r\frac{\d z_i^+}{2\pi\i z_i^+}\prod_{j=1}^c\frac{\d z_j^-}{2\pi\i z_j^-}\Big(\prod_{i}  z_i^+\Big)^{-\zeta_+}\Big(\prod_{j}  z_j^-\Big)^{\zeta_-}\times\\
\times\Delta(z^+,z^-;\q,\t,\p)\, \prod_{i=1}^r\frac{(z^+_i/\alpha_+\mu_-\p^{-2};\q)_\infty}{(z^+_i/\alpha_+\mu_+;\q)_\infty}\prod_{j=1}^c\frac{(z^-_j/\alpha_-\mu_+\p^2;\t)_\infty}{(z^-_j/\alpha_-\mu_-;\t)_\infty}~,
\end{multline}
provided $\alpha_+/\alpha_-=\p^{3/2}$, $\p^{-1}\q^{\zeta_+}=\Lambda=\p^{-1}\t^{\zeta_-}$. Comparing with the defect partition function (\ref{eq:DefectMMqtinv}) coming from the ordinary SQED, the two are related by analytic continuation $\t^{-1}\to \t$ and removing boundary contributions (Theta functions/$\t$-constants from the measure), together with straightforward identification of parameters. In particular, the contours are different because the pole structure is different, and in the supergroup case there are no poles from the would be intersection sector.\\

\noindent\textbf{Single Component $\text{U}(r|0)\times \text{U}(r'|0)$ Defect.} We are here interested in
\be
\nu_+\to \nu_+^*\equiv \mu_+\t^{-r}~,\quad \nu_-\to \nu_-^*\equiv \mu_-\p^{-1}\t^{r'}~,
\ee
so that $\ell(\lambda^+)\leq r$, $\ell(\lambda^{-})\leq r'$. In this case, it is convenient to write the bi-fundamental contributions in the equivalent infinite product form
\bes
\begin{align}
Z^\text{v}_{+-} = & \ \prod_{(x,x')\in\chi^+\times\chi^-}\frac{(\t x/x';\q)_\infty}{(\t^2 x/x';\q)_\infty}\prod_{(x,x')\in\chi^+_\emptyset\times\chi^-_\emptyset}\frac{(\t^2 x/x';\q)_\infty}{(\t x/x';\q)_\infty}~,\\
Z^\text{v}_{-+} = & \ \prod_{(x,x')\in\chi^+\times\chi^-}\frac{(\t^{-1} x'/x;\q)_\infty}{(x'/x;\q)_\infty}\prod_{(x,x')\in\chi^+_\emptyset\times\chi^-_\emptyset}\frac{(x'/x;\q)_\infty}{(\t^{-1}x'/x;\q)_\infty}~,
\end{align}
\ees
while for the diagonal terms we use $Z_{--}^\text{v}=Z_{++}^\text{v}|_{\lambda^+\to \lambda^-}$. Note that we introduced the set
\be
\chi^-\equiv \{\eta_-\nu_-\t^{-(i-1)}\q^{\lambda^{-}_i},\lambda^{-}_i\geq \lambda^{-}_{i+1},\i\in[1,+\infty)\}~,
\ee
simply related to $\chi^+$ by $(\pm,\q,\t)\to (\mp,\q^{-1},\t^{-1})$: this is why the $-$ sector is associated to a negative gauge node. At the specified points, we get the partial simplification in the $--$ sector
\be
Z_{--}^\text{v}Z_{--}^\text{af}\!\!\!\xrightarrow{\nu_-\to \nu_-^*}  \!\!\!\prod_{i,j=1}^{r'}\frac{(\t\,\t^{j-i};\q)_\infty}{(\t^{j-i};\q)_\infty}\frac{(\t^{j-i}\q^{\lambda^-_i-\lambda^-_j};\q)_\infty}{(\t\, \t^{j-i}\q^{\lambda^-_i-\lambda^-_j};\q)_\infty}\frac{(\p^{1/2}\t^{-r'})^{|\lambda^-|}}{N_{ \emptyset \lambda^-}(\t^{-r'};\q,\t)f_{\lambda^-}(\q,\t)}~,
\ee
while in the $++$ the simplification is as before. Eventually, the truncated super-instanton summands can be organized as the (normalized) residues at the poles
\be
z^\pm_i=\alpha_\pm\mu_\pm \t^{\pm(i-r)}\q^{\mp\lambda_i^\pm}~,
\ee
in the contour integral 
\begin{multline}
Z_\text{inst.}[\text{U}(1|1)+\text{af}]\!\xrightarrow{\nu_\pm\to \nu_\pm^*}\!\!\oint\prod_{i=1}^r\frac{\d z_i^+}{2\pi\i z_i^+}\prod_{j=1}^{r'}\frac{\d z_j^-}{2\pi\i z_j^-}\Big(\prod_{i}  z_i^+\Big)^{-\zeta_+}\Big(\prod_{j}  z_j^-\Big)^{-\zeta_-}\times\\
\times\Delta_\t(z^+;\q)\Delta_\t(z^-;\q)\, \prod_{i=1}^r\prod_{j=1}^{r'}\frac{(z^+_i/z^-_j v;\q)_\infty}{(\t z^+_i/z^-_j v;\q)_\infty}\frac{(v z^-_j/z^+_i;\q)_\infty}{(\t v z^-_j/z^+_i;\q)_\infty}\times\\
\times \prod_{i=1}^r\frac{(\t \alpha_+\mu_-/z^+_i;\q)_\infty}{(z^+_i/\alpha_+\mu_+;\q)_\infty}\prod_{i=1}^{r'}\frac{(\t z^-_i/\alpha_-\mu_+;\q)_\infty}{(\alpha_-\mu_-/z^-_i;\q)_\infty}
\end{multline}
provided  $\q^{\zeta_+}=\Lambda=\q^{\zeta_-}\mu_-/\mu_+$, where we also set $v\equiv \t \p^{-1}\alpha_+/\alpha_-$ (this shift may be reabsorbed). The resulting measure looks like that of $\textrm{U}(r+r')$ refined Chern-Simons broken to $\textrm{U}(r)\times\textrm{U}(r')$ by the potential. This is very reminiscent of the lens space $L(2,1)$ matrix model (with eigenvalues placed around two distinct connections), however, we do not currently have an interpretation in this direction.

\section{Summary and Discussion}
Motivated by a common algebraic engineering origin, we considered two a priori distinct matrix-like models: upon localization, one is associated to a coupled 3d-1d intersecting defect theory arising from Higgsing a parent 5d $\text{U}(1)$ theory, the other one arises from Higgsing a 5d $\text{U}(1|1)$ theory. It turns out that the two are essentially related by analytic continuation in one of the equivariant parameters. This seems to parallel what is known from other situations \cite{Marino:2009jd}: the $\mathbb{S}^3$ $\textrm{U}(N_1)\times \textrm{U}(N_2)$ ABJ(M) and the $L(2,1)$ $\textrm{U}(N_1+N_2)$ Chern-Simons matrix models are related by analytic continuation $N_2\to -N_2$, while from the combinatorial perspective the role of matrix sizes is played by $1/\ln\q$, $-1/\ln\t$ \cite{Eynard:2015aea}. Furthermore, we considered yet another Higgsing of the 5d $\text{U}(1|1)$ theory, and the partition function of the resulting single component defect theory resembles that of refined Chern-Simons on $L(2,1)$. 

It is well-known that the large rank expansion of a matrix model and its supergroup version are equivalent up to non-perturbative effects \cite{Vafa:2014iua} (in the unrefined case, this fits with the generic Higgsing being sensitive only to $r-c$): it would be interesting to retrace and adapt the existing analysis around supergroups, large rank dualities and non-perturbative effects in our refined setup, also in view of the vertex/anti-vertex formalism \cite{Kimura:2020lmc}. The relations between supergroup-like and ordinary theories was instrumental for understanding non-perturbative effects in topological strings \cite{Hatsuda:2013oxa}, most notably on the local $\mathbb{P}^1\times\mathbb{P}^1$ geometry, the closed dual to Chern-Simons on $L(2,1)$: a deeper understanding of the subject reviewed in this note may help in shedding more light on seemingly different proposals \cite{Lockhart:2012vp}. On the more mathematical side, it would be interesting to study the moduli space of generalized defects and their relations to the parent instanton ones, as well as to explore the possible extension to the intersecting/supergroup setup intriguing dualities between quivers, knots and Donaldson-Thomas invariants \cite{Kucharski:2017poe}.

\begin{acknowledgement}
The work of FN is supported by the Simons Bridge Fellowship and the Alexander von Humboldt Foundation.
\end{acknowledgement}

\section*{Appendix}\label{app:conventions}

\addcontentsline{toc}{section}{Appendix}

We summarize the definitions and some property of the special functions we use throughout the main text. The (infinite) $\q$-Pochhammer symbol or $\q$-factorial is defined by 
\be
(x;\q)_\infty\equiv\prod_{k\geq 0}(1-\q^k x)~,\quad |\q|<1~,
\ee
and it can be extended to $|\q|>1$ by means of $(\q x;\q)_\infty\to \frac{1}{(x;\q^{-1})_\infty}$. The short Jacobi Theta function is defined by
\be\label{Theta}
\Theta(x;\q)\equiv (x;\q)_\infty(\q x^{-1};\q)_\infty~.
\ee

Nekrasov's function is defined according to
\be
N_{\mu\nu}(x;\q,\t)\equiv \prod_{(i,j)\in\mu}(1-x \q^{\mu_i-j}\t^{\nu^\vee_j-i+1})\prod_{(i,j)\in\nu}(1-x \q^{-\nu_i+j-1}\t^{-\mu^\vee_j+i})~,
\ee
where $\mu$, $\nu$ are integer partitions or Young diagrams (i.e. $\mu_{i}\geq \mu_{i+1}\geq 0$, $\nu_{i}\geq \nu_{i+1}\geq 0$) parametrized by the coordinates $(i,j)$ of boxes running over the rows and columns respectively, with ${}^\vee$ the transpose operation. Its properties and diverse representations can be found in \cite{Awata:2008ed}.
%
%
%


\begin{thebibliography}{10}

\bibitem{Aharony:2008ug}
O.~Aharony, O.~Bergman, D.~L. Jafferis, and J.~Maldacena, ``{N=6 superconformal
  Chern-Simons-matter theories, M2-branes and their gravity duals},''
  \href{http://dx.doi.org/10.1088/1126-6708/2008/10/091}{{\em JHEP} {\bf 10}
  (2008)  091}, \href{http://arxiv.org/abs/0806.1218}{{\tt arXiv:0806.1218
  [hep-th]}}.

\bibitem{Aharony:2008gk}
O.~Aharony, O.~Bergman, and D.~L. Jafferis, ``{Fractional M2-branes},''
  \href{http://dx.doi.org/10.1088/1126-6708/2008/11/043}{{\em JHEP} {\bf 11}
  (2008)  043}, \href{http://arxiv.org/abs/0807.4924}{{\tt arXiv:0807.4924
  [hep-th]}}.

\bibitem{Kapustin:2009kz}
A.~Kapustin, B.~Willett, and I.~Yaakov, ``{Exact Results for Wilson Loops in
  Superconformal Chern-Simons Theories with Matter},''
  \href{http://dx.doi.org/10.1007/JHEP03(2010)089}{{\em JHEP} {\bf 03} (2010)
  089}, \href{http://arxiv.org/abs/0909.4559}{{\tt arXiv:0909.4559 [hep-th]}}.

\bibitem{Drukker:2009hy}
N.~Drukker and D.~Trancanelli, ``{A Supermatrix model for N=6 super
  Chern-Simons-matter theory},''
  \href{http://dx.doi.org/10.1007/JHEP02(2010)058}{{\em JHEP} {\bf 02} (2010)
  058}, \href{http://arxiv.org/abs/0912.3006}{{\tt arXiv:0912.3006 [hep-th]}}.

\bibitem{Marino:2009jd}
M.~Marino and P.~Putrov, ``{Exact Results in ABJM Theory from Topological
  Strings},'' \href{http://dx.doi.org/10.1007/JHEP06(2010)011}{{\em JHEP} {\bf
  06} (2010)  011}, \href{http://arxiv.org/abs/0912.3074}{{\tt arXiv:0912.3074
  [hep-th]}}.

\bibitem{Awata:2012jb}
H.~Awata, S.~Hirano, and M.~Shigemori, ``{The Partition Function of ABJ
  Theory},'' \href{http://dx.doi.org/10.1093/ptep/ptt014}{{\em PTEP} {\bf 2013}
  (2013)  053B04}, \href{http://arxiv.org/abs/1212.2966}{{\tt arXiv:1212.2966
  [hep-th]}}.

\bibitem{Drukker:2010nc}
N.~Drukker, M.~Marino, and P.~Putrov, ``{From weak to strong coupling in ABJM
  theory},'' \href{http://dx.doi.org/10.1007/s00220-011-1253-6}{{\em Commun.
  Math. Phys.} {\bf 306} (2011)  511--563},
  \href{http://arxiv.org/abs/1007.3837}{{\tt arXiv:1007.3837 [hep-th]}}.

\bibitem{Drukker:2011zy}
N.~Drukker, M.~Marino, and P.~Putrov, ``{Nonperturbative aspects of ABJM
  theory},'' \href{http://dx.doi.org/10.1007/JHEP11(2011)141}{{\em JHEP} {\bf
  11} (2011)  141}, \href{http://arxiv.org/abs/1103.4844}{{\tt arXiv:1103.4844
  [hep-th]}}.

\bibitem{Marino:2011eh}
M.~Marino and P.~Putrov, ``{ABJM theory as a Fermi gas},''
  \href{http://dx.doi.org/10.1088/1742-5468/2012/03/P03001}{{\em J. Stat.
  Mech.} {\bf 1203} (2012)  P03001}, \href{http://arxiv.org/abs/1110.4066}{{\tt
  arXiv:1110.4066 [hep-th]}}.

\bibitem{Hatsuda:2013oxa}
Y.~Hatsuda, M.~Marino, S.~Moriyama, and K.~Okuyama, ``{Non-perturbative effects
  and the refined topological string},''
  \href{http://dx.doi.org/10.1007/JHEP09(2014)168}{{\em JHEP} {\bf 09} (2014)
  168}, \href{http://arxiv.org/abs/1306.1734}{{\tt arXiv:1306.1734 [hep-th]}}.

\bibitem{Vafa:2014iua}
C.~Vafa, ``{Non-Unitary Holography},''
  \href{http://arxiv.org/abs/1409.1603}{{\tt arXiv:1409.1603 [hep-th]}}.

\bibitem{Dijkgraaf:2016lym}
R.~Dijkgraaf, B.~Heidenreich, P.~Jefferson, and C.~Vafa, ``{Negative Branes,
  Supergroups and the Signature of Spacetime},''
  \href{http://dx.doi.org/10.1007/JHEP02(2018)050}{{\em JHEP} {\bf 02} (2018)
  050}, \href{http://arxiv.org/abs/1603.05665}{{\tt arXiv:1603.05665
  [hep-th]}}.

\bibitem{Kimura:2019msw}
T.~Kimura and V.~Pestun, ``{Super instanton counting and localization},''
  \href{http://arxiv.org/abs/1905.01513}{{\tt arXiv:1905.01513 [hep-th]}}.

\bibitem{Chen:2019vvt}
H.-Y. Chen, T.~Kimura, and N.~Lee, ``{Quantum Elliptic Calogero-Moser Systems
  from Gauge Origami},'' \href{http://dx.doi.org/10.1007/JHEP02(2020)108}{{\em
  JHEP} {\bf 02} (2020)  108}, \href{http://arxiv.org/abs/1908.04928}{{\tt
  arXiv:1908.04928 [hep-th]}}.

\bibitem{Chen:2020rxu}
H.-Y. Chen, T.~Kimura, and N.~Lee, ``{Quantum Integrable Systems from
  Supergroup Gauge Theories},''
  \href{http://dx.doi.org/10.1007/JHEP09(2020)104}{{\em JHEP} {\bf 09} (2020)
  104}, \href{http://arxiv.org/abs/2003.13514}{{\tt arXiv:2003.13514
  [hep-th]}}.

\bibitem{Nekrasov:2018gne}
N.~Nekrasov, ``{Superspin chains and supersymmetric gauge theories},''
  \href{http://dx.doi.org/10.1007/JHEP03(2019)102}{{\em JHEP} {\bf 03} (2019)
  102}, \href{http://arxiv.org/abs/1811.04278}{{\tt arXiv:1811.04278
  [hep-th]}}.

\bibitem{Kimura_2021}
T.~Kimura and F.~Nieri,
  ``{Intersecting defects and
  supergroup gauge theory},'' \href{http://dx.doi.org/10.1088/1751-8121/ac2716}{{\em Journal of Physics A: Mathematical and
  Theoretical} {\bf 54} (sep, 2021)  435401}.

\bibitem{Aganagic:2011sg}
M.~Aganagic and S.~Shakirov, ``{Knot Homology and Refined Chern-Simons
  Index},'' \href{http://dx.doi.org/10.1007/s00220-014-2197-4}{{\em Commun.
  Math. Phys.} {\bf 333} (2015) no.~1, 187--228},
  \href{http://arxiv.org/abs/1105.5117}{{\tt arXiv:1105.5117 [hep-th]}}.

\bibitem{Cassia:2021uly}
L.~Cassia and M.~Zabzine, ``On refined Chern-Simons and refined ABJ matrix models,''
[arXiv:2107.07525 [hep-th]].

\bibitem{Nekrasov:2002qd}
N.~A. Nekrasov, ``{Seiberg-Witten prepotential from instanton counting},''
  \href{http://dx.doi.org/10.4310/ATMP.2003.v7.n5.a4}{{\em Adv. Theor. Math.
  Phys.} {\bf 7} (2003) no.~5, 831--864},
  \href{http://arxiv.org/abs/hep-th/0206161}{{\tt arXiv:hep-th/0206161}}.

\bibitem{Nekrasov:2003rj}
N.~Nekrasov and A.~Okounkov, ``{Seiberg-Witten theory and random partitions},''
  \href{http://dx.doi.org/10.1007/0-8176-4467-9_15}{{\em Prog. Math.} {\bf 244}
  (2006)  525--596}, \href{http://arxiv.org/abs/hep-th/0306238}{{\tt
  arXiv:hep-th/0306238}}.

\bibitem{Gomis:2016ljm}
J.~Gomis, B.~Le~Floch, Y.~Pan, and W.~Peelaers, ``{Intersecting Surface Defects
  and Two-Dimensional CFT},''
  \href{http://dx.doi.org/10.1103/PhysRevD.96.045003}{{\em Phys. Rev. D} {\bf
  96} (2017) no.~4, 045003}, \href{http://arxiv.org/abs/1610.03501}{{\tt
  arXiv:1610.03501 [hep-th]}}.

\bibitem{Pan:2016fbl}
Y.~Pan and W.~Peelaers, ``{Intersecting Surface Defects and Instanton Partition
  Functions},'' \href{http://dx.doi.org/10.1007/JHEP07(2017)073}{{\em JHEP}
  {\bf 07} (2017)  073}, \href{http://arxiv.org/abs/1612.04839}{{\tt
  arXiv:1612.04839 [hep-th]}}.

\bibitem{Nieri:2017ntx}
F.~Nieri, Y.~Pan, and M.~Zabzine, ``{3d Expansions of 5d Instanton Partition
  Functions},'' \href{http://dx.doi.org/10.1007/JHEP04(2018)092}{{\em JHEP}
  {\bf 04} (2018)  092}, \href{http://arxiv.org/abs/1711.06150}{{\tt
  arXiv:1711.06150 [hep-th]}}.

\bibitem{Nieri:2018pev}
F.~Nieri, Y.~Pan, and M.~Zabzine, ``{3d Mirror Symmetry from S-duality},''
  \href{http://dx.doi.org/10.1103/PhysRevD.98.126002}{{\em Phys. Rev. D} {\bf
  98} (2018) no.~12, 126002}, \href{http://arxiv.org/abs/1809.00736}{{\tt
  arXiv:1809.00736 [hep-th]}}.

\bibitem{sergeev2008deformed}
A.~N. Sergeev and A.~P. Veselov, ``Deformed macdonald-ruijsenaars operators and
  super macdonald polynomials,'' Communications in Mathematical Physics 288.2 (2009): 653-675, \href{http://arxiv.org/abs/0707.3129}{{\tt
  arXiv:0707.3129 [math.QA]}}.
 

\bibitem{atai2021supermacdonald}
F.~Atai, M.~Hallnäs, and E.~Langmann, ``Super-macdonald polynomials:
  Orthogonality and Hilbert space interpretation,'' \href{http://dx.doi.org/10.1007/s00220-021-04166-z}{Communications in Mathematical Physics, Volume 388, Issue 1, p.435-468},  \href{http://arxiv.org/abs/2103.07400}{{\tt arXiv:2103.07400 [math.QA]}}.

\bibitem{Atai_2019}
F.~{Atai}, M.~{Halln{\"a}s}, and E.~{Langmann}, ``{Orthogonality of super-Jack
  polynomials and a Hilbert space interpretation of deformed
  Calogero-Moser-Sutherland operators},'' \href{http://dx.doi.org/10.1112/blms.12234}{Bulletin of the London Mathematical Society 51.2 (2019): 353-370},
  \href{http://arxiv.org/abs/1802.02016}{{\tt arXiv:1802.02016 [math.QA]}}.

\bibitem{Vafa:2001qf}
C.~Vafa, ``{Brane / anti-brane systems and U(N|M) supergroup},''
  \href{http://arxiv.org/abs/hep-th/0101218}{{\tt arXiv:hep-th/0101218}}.

\bibitem{Okuda:2006fb}
T.~Okuda and T.~Takayanagi, ``{Ghost D-branes},''
  \href{http://dx.doi.org/10.1088/1126-6708/2006/03/062}{{\em JHEP} {\bf 03}
  (2006)  062}, \href{http://arxiv.org/abs/hep-th/0601024}{{\tt
  arXiv:hep-th/0601024}}.

\bibitem{Mikhaylov:2014aoa}
V.~Mikhaylov and E.~Witten, ``{Branes And Supergroups},''
  \href{http://dx.doi.org/10.1007/s00220-015-2449-y}{{\em Commun. Math. Phys.}
  {\bf 340} (2015) no.~2, 699--832}, \href{http://arxiv.org/abs/1410.1175}{{\tt
  arXiv:1410.1175 [hep-th]}}.

\bibitem{Gopakumar:1998ki}
R.~Gopakumar and C.~Vafa, ``{On the gauge theory / geometry correspondence},''
  \href{http://dx.doi.org/10.4310/ATMP.1999.v3.n5.a5}{{\em Adv. Theor. Math.
  Phys.} {\bf 3} (1999)  1415--1443},
  \href{http://arxiv.org/abs/hep-th/9811131}{{\tt arXiv:hep-th/9811131}}.

\bibitem{Kimura:2020jxl}
T.~Kimura, \href{http://dx.doi.org/10.1007/978-3-030-76190-5}{{\em {Instanton
  Counting, Quantum Geometry and Algebra}}}.
\newblock Springer, 7, 2021.
\newblock \href{http://arxiv.org/abs/2012.11711}{{\tt arXiv:2012.11711
  [hep-th]}}.

\bibitem{Awata:2016riz}
H.~Awata, H.~Kanno, T.~Matsumoto, A.~Mironov, A.~Morozov, A.~Morozov,
  Y.~Ohkubo, and Y.~Zenkevich, ``{Explicit examples of DIM constraints for
  network matrix models},''
  \href{http://dx.doi.org/10.1007/JHEP07(2016)103}{{\em JHEP} {\bf 07} (2016)
  103}, \href{http://arxiv.org/abs/1604.08366}{{\tt arXiv:1604.08366
  [hep-th]}}.

\bibitem{Zenkevich:2018fzl}
Y.~Zenkevich, ``{Higgsed network calculus},''
  \href{http://arxiv.org/abs/1812.11961}{{\tt arXiv:1812.11961 [hep-th]}}.

\bibitem{Shiraishi:1995rp}
J.~Shiraishi, H.~Kubo, H.~Awata, and S.~Odake, ``{A Quantum deformation of the
  Virasoro algebra and the Macdonald symmetric functions},''
  \href{http://dx.doi.org/10.1007/BF00398297}{{\em Lett.Math.Phys.} {\bf 38}
  (1996)  33--51},
\href{http://arxiv.org/abs/q-alg/9507034}{{\tt arXiv:q-alg/9507034 [q-alg]}}.

\bibitem{frenkelR}
E.~Frenkel and N.~Reshetikhin, ``{Quantum Affine Algebras and Deformations of
  the Virasoro and W-algebras},''
  \href{http://dx.doi.org/10.10072FBF02104917}{{\em Commun.Math.Phys.} {\bf
  178} (1996)  237--264}, \href{http://arxiv.org/abs/q-alg/9505025}{{\tt
  arXiv:q-alg/9505025 [q-alg]}}.

\bibitem{Aganagic:2013tta}
M.~Aganagic, N.~Haouzi, C.~Kozcaz, and S.~Shakirov, ``{Gauge/Liouville
  Triality},'' \href{http://arxiv.org/abs/1309.1687}{{\tt arXiv:1309.1687
  [hep-th]}}.

\bibitem{Kimura:2015rgi}
T.~Kimura and V.~Pestun, ``{Quiver W-algebras},''
  \href{http://dx.doi.org/10.1007/s11005-018-1072-1}{{\em Lett. Math. Phys.}
  {\bf 108} (2018) no.~6, 1351--1381},
  \href{http://arxiv.org/abs/1512.08533}{{\tt arXiv:1512.08533 [hep-th]}}.
  
\bibitem{Chen:2020rxu}
H.~Y.~Chen, T.~Kimura and N.~Lee, ``{Quantum Integrable Systems from Supergroup Gauge Theories},"
\href{http://dx.doi.org/10.1007/JHEP09(2020)104}{{\em JHEP09(2020)104}}, \href{http://arxiv.org/abs/2003.13514}{{\tt arXiv:arXiv:2003.13514 [hep-th]}}.


\bibitem{Eynard:2015aea}
B.~Eynard, T.~Kimura, and S.~Ribault, ``{Random matrices},''
  \href{http://arxiv.org/abs/1510.04430}{{\tt arXiv:1510.04430 [math-ph]}}.

\bibitem{Kimura:2020lmc}
T.~Kimura and Y.~Sugimoto, ``{Topological Vertex/anti-Vertex and Supergroup
  Gauge Theory},'' \href{http://dx.doi.org/10.1007/JHEP04(2020)081}{{\em JHEP}
  {\bf 04} (2020)  081}, \href{http://arxiv.org/abs/2001.05735}{{\tt
  arXiv:2001.05735 [hep-th]}}.

\bibitem{Lockhart:2012vp}
G.~Lockhart and C.~Vafa, ``{Superconformal Partition Functions and
  Non-perturbative Topological Strings},''
  \href{http://dx.doi.org/10.1007/JHEP10(2018)051}{{\em JHEP} {\bf 10} (2018)
  051}, \href{http://arxiv.org/abs/1210.5909}{{\tt arXiv:1210.5909 [hep-th]}}.

\bibitem{Kucharski:2017poe}
P.~Kucharski, M.~Reineke, M.~Stosic, and P.~Su\l{}kowski, ``{BPS states, knots
  and quivers},'' \href{http://dx.doi.org/10.1103/PhysRevD.96.121902}{{\em
  Phys. Rev. D} {\bf 96} (2017) no.~12, 121902},
  \href{http://arxiv.org/abs/1707.02991}{{\tt arXiv:1707.02991 [hep-th]}}.

\bibitem{Awata:2008ed}
H.~Awata and H.~Kanno, ``{Refined BPS state counting from Nekrasov's formula
  and Macdonald functions},''
  \href{http://dx.doi.org/10.1142/S0217751X09043006}{{\em Int. J. Mod. Phys. A}
  {\bf 24} (2009)  2253--2306}, \href{http://arxiv.org/abs/0805.0191}{{\tt
  arXiv:0805.0191 [hep-th]}}.

\end{thebibliography}
%

\providecommand{\href}[2]{#2}\begingroup\raggedright\endgroup

\end{document}